\documentclass[12pt]{article}
\usepackage{latexsym}
\usepackage{amssymb}
\usepackage{array}

\def\0{\over } \def\1{\vec } \def\2{{1\over2}} \def\4{{1\over4}}
\def\5{\bar } %\def\5{\overline }
\def\6{\partial }
\def\7#1{{#1}\llap{/}}
\def\8#1{{\textstyle{#1}}} \def\9#1{{\bf{#1}}}
\def\.{\cdot }
\def\^#1{\widehat{#1}}

\def\({\left(} \def\){\right)} \def\<{\langle } \def\>{\rangle }
\def\[{\left[} \def\]{\right]}  
 
\def\pmbf#1{\setbox0=\hbox{${#1}$}
        \kern-.025em\copy0\kern-\wd0
        \kern.05em\copy0\kern-\wd0
        \kern-.025em\raise.0433em\box0 }

\def\be{\begin{equation}}
\def\ee{\end{equation}}

\newcommand{\bel}[1]{\begin{equation}\label{#1}}
\def\bea{\begin{eqnarray}}
\newcommand{\beal}[1]{\begin{eqnarray}\label{#1}}
\def\eea{\end{eqnarray}}

\newcommand{\unit}{1\!\!1}
\newcommand{\intsum}{\sum\!\!\!\!\!\!\!\int}  

\begin{document}

\begin{titlepage}
\renewcommand{\thefootnote}{\alph{footnote}}
~\vspace{-2cm}
\begin{flushright} 
TUW-04-02,
YITP-04-03, ITP-UH-02/04
\end{flushright}  
%\begin{center}  \vfil %\vspace{1.6cm}
%{%\large 
\vfil
\centerline{\Large\bf A new anomalous contribution to 
%the central charge of the N=2 monopole
}
\medskip
\centerline{\Large\bf %A new anomalous contribution to 
the central charge of the N=2 monopole}
%\medskip
%\centerline{\Large\bf and one-loop mass corrections}
%\medskip
%\centerline{\Large\bf A central charge anomaly in the N=2 monopole}
\medskip
\begin{center}
\vfil 
{\large  
A. Rebhan$^1$\footnote{\footnotesize\tt rebhana@hep.itp.tuwien.ac.at}, 
P. van Nieuwenhuizen$^2$\footnote{\footnotesize\tt vannieu@insti.physics.sunysb.edu} 
and R. Wimmer$^3$\footnote{\footnotesize\tt wimmer@itp.uni-hannover.de}
}\\
\end{center}  \medskip \smallskip \qquad \qquad 
{\sl $^1$} \parbox[t]{12cm}{\sl 
  Institut f\"ur Theoretische Physik, Technische Universit\"at Wien, \\
  Wiedner Hauptstr. 8--10, A-1040 Vienna, Austria\\ } \\
\bigskip \qquad \qquad 
{\sl $^2$} \parbox[t]{12cm}{\sl 
  C.N.Yang Institute for Theoretical Physics, \\
  SUNY at Stony Brook, Stony Brook, NY 11794-3840, USA } 
\medskip \qquad \qquad 
{\sl $^3$} \parbox[t]{12cm}{\sl 
Institut f\"ur Theoretische Physik, Universit\"at Hannover,\\ 
Appelstr.~2, D-30167 Hanover, Germany\\ } \\
\vfil
\centerline{ABSTRACT}\vspace{.5cm}
We calculate the one-loop corrections to the mass and central charge
of the BPS monopole in $N=2$ super-Yang-Mills theory in 3+1 dimensions
using a supersymmetry-preserving version of dimensional regularization
{adapted to solitons}.
In the renormalization scheme where previous studies have indicated
vanishing quantum corrections,
we find nontrivial corrections that we identify as
the 3+1 dimensional analogue of the {anomaly
in the conformal central charge of the}
$N=1$ supersymmetric kink in 1+1 dimensions.
As in the latter case, the {associated}
%anomalous 
contribution to the {ordinary} central charge has
exactly the required magnitude to preserve
BPS saturation at the one-loop level.
%In the case of the $N=2$ monopole, i
It also restores consistency
of calculations involving sums over zero-point energies
with the low-energy effective action of Seiberg and Witten.

\vfil
\end{titlepage}

\setcounter{footnote}{0}

Supersymmetric (susy) solitons which saturate the Bogomolnyi bound 
\cite{Bogomolny:1976de,Witten:1978mh}
have been found in various models and in various dimensions, and they
play an important role in recent studies of nonperturbative
effects in (susy) field theory and of duality
\cite{Seiberg}.

In the earliest calculation of quantum corrections to these
solitons it was assumed that supersymmetry
would ensure complete cancellation of quantum corrections \cite{D'Adda:1978ur},
thereby trivially guaranteeing Bogomolnyi-Prasad-Sommerfield 
(BPS) saturation \cite{Bogomolny:1976de,Prasad:1975kr} 
at the quantum level. However, more careful calculations
found that there may be nonzero corrections already
in the simplest example of the 1+1-dimensional susy kink,
but by the end of the 1980's
the literature was in unresolved disagreement concerning the
correct value. This issue was reopened in 1997 when two of
us \cite{Rebhan:1997iv} noted that some of the earlier calculations had used
methods which gave wrong results when applied to the exactly
solvable bosonic sine-Gordon kink.
The result for the mass obtained in \cite{Rebhan:1997iv}
was however contaminated by energy located at the boundary of
the quantization volume, which was corrected subsequently
in \cite{Nastase:1998sy} by the use of topological boundary
conditions. This singled out as correct the earlier result
of Schonfeld \cite{Schonfeld:1979hg}, who considered a kink-antikink
system,
as well as of Casahorr{\'a}n~\cite{Casahorran:1989vd}, who
used a finite mass formula in terms of only
the discrete modes \cite{Cahill:1976im}, 
and refuted the null results of 
Refs.~\cite{Kaul:1983yt,Chatterjee:1984xh,Yamagishi:1984zv,Uchiyama:1984kb}.
However it led to a new problem
because it seemed to imply a violation of the Bogomolnyi bound,
since the central charge did not appear to receive quantum
corrections \cite{Imbimbo:1984nq}.

Ref.~\cite{Nastase:1998sy} suspected an anomaly at work,
{and shortly thereafter Shifman, Vainshtein, and Voloshin
\cite{Shifman:1998zy} showed that supersymmetry enforces
an anomalous contribution to the central charge which} is
%and an anomaly in the central charge which
%restores BPS saturation was
%shortly thereafter derived by Shifman, Vainshtein, and Voloshin
%\cite{Shifman:1998zy}. The central charge anomaly is 
in fact
part of a susy multiplet involving other better-known anomalies,
the trace and conformal-susy anomaly.

In Ref.~\cite{Rebhan:2002uk} we recently developed a version
of dimensional regularization which preserves susy
and reproduces the correct susy kink mass without
the complications of other methods. In Ref.~\cite{Rebhan:2002yw}
we then demonstrated how the {anomalous contribution to} the central charge
of the susy kink can be obtained as a remnant of parity violation
in the odd-dimensional model used for embedding the susy kink.
{We also showed that the anomaly in question is not an anomaly in
the ordinary central charge, but in the conformal central charge.
The ordinary central charge itself has no anomaly, because it is
produced by the anticommutator of two ordinary susy charges and the latter
are free from anomalies. However, we found that one can define a
conformal central-charge current whose divergence is proportional
to the ordinary central-charge current, and the anomaly in the
former is directly related to nontrivial quantum corrections to the latter.
}

Most recently, we applied
this method to the $N=2$ vortex in 2+1 dimensions \cite{Rebhan:2003bu}
and demonstrated
BPS saturation despite nonvanishing quantum corrections. However,
in this case these quantum corrections are 
{in no way related to an anomaly (in the conformal central charge),
as there is also no trace and conformal-susy anomaly in odd dimensions.
Nevertheless, the corrections to the central charge 
%are entirely non-anomalous, albeit 
turned out to be}
nontrivial in that
one has to take into account the effect of the background vortex
on the quantum fluctuations far away from the vortex.
The value of the mass correction agreed with a result
deduced by heat-kernel methods \cite{Vassilevich:2003xk}.

In this Letter, we consider the $N=2$ monopole in 3+1 dimensions,
which has been used by many authors in studies of duality.
The monopole model has more unbroken susy generators than the
susy kink or the vortex, so one runs the risk (or the blessing) of
vanishing quantum corrections. This model has been studied
before in Refs.~\cite{D'Adda:1978mu,Kaul:1984bp,Imbimbo:1985mt}
and while the initial result of vanishing corrections of
Ref.~\cite{D'Adda:1978mu} turned out to be an oversimplification,
Refs.~\cite{Kaul:1984bp,Imbimbo:1985mt} nevertheless arrived at
the conclusion of vanishing quantum corrections, at least in
the simplest renormalization scheme.

{By setting up a suitable susy-preserving dimensional regularization
method which embeds the monopole into 4+1 dimensions,
we shall verify explicitly that BPS saturation holds\footnote{%
Since BPS saturation is guaranteed by the multiplet shortening arguments of 
Ref.~\cite{Witten:1978mh}, this just verifies that our regularization method
indeed preserves susy.}, but we find a nonvanishing
contribution to the mass (in the simplest renormalization scheme),
which is matched by an anomalous
contribution to the central charge operator coming 
from parity-violating quantum corrections to
the additional momentum component, precisely analogous
to the situation in the susy kink.
}

%\section{Embedding the {\cal{N}}=2 monopole}

The $N=2$ super-Yang-Mills theory in 3+1 dimensions can be obtained
by dimensional reduction from the (5+1)-dimensional $N=1$ theory
%The Lagrangian for the six-dimensional super-YM theory is given by 
\cite{Brink:1977bc} 
\begin{equation}
  \label{eq:L6d}
  \mathcal L
=- \frac{1}{4} F_{AB}^2 - \bar{\lambda}\Gamma^AD_A\lambda,
\end{equation}
where the indices $A,B$ take the values $0,1,2,3,5,6$
and which is invariant under
\be
\delta A_B^a=\bar\lambda^a \Gamma_B \eta-\bar\eta \Gamma_B \lambda^a ,\quad
\delta \lambda^a={1\02} F^a_{BC}\Gamma^B \Gamma^C \eta.\ee
The complex spinor 
$\lambda$ is in the adjoint representation of the gauge group
which we assume to be SU(2) in the following and 
$(D_A\lambda)^a=(\partial_A\lambda+gA_A\times \lambda)^a\break
=\partial_A\lambda^a+g\epsilon^{abc}A_A^b\lambda^c$. 
Furthermore, $\lambda$ and $\eta$ satisfy the Weyl condition:
\begin{equation}
  \label{eq:wcon}
  (1-\Gamma_7)\lambda=0\quad \textrm{with} \quad 
  \Gamma_7=\Gamma_0\Gamma_1\Gamma_2\Gamma_3\Gamma_5\Gamma_6. 
\end{equation}

To carry out the dimensional reduction we write
%Dimensional reduction 
$A_B=(A_\mu,P,S)$ and choose the following representation of
gamma matrices
\begin{eqnarray}
  \label{eq:6dgam}
  \Gamma_\mu=\gamma_\mu\otimes\sigma_1\ &,&\ \mu=0,1,2,3,\nonumber\\
  \Gamma_5=\gamma_5\otimes\sigma_1\ &,&\ \Gamma_6=\unit\otimes\sigma_2.
\end{eqnarray}
In this representation
the Weyl condition (\ref{eq:wcon}) becomes 
$\lambda={\psi\choose 0}$, 
with a complex four-component spinor $\psi$.\footnote{We 
use the metric with signature $(-,+,+,+,+,+)$ and 
$\bar\lambda^a=(\lambda^a)^\dagger i \Gamma^0$,
hence $\bar\psi^a=(\psi^a)^\dagger i \gamma^0$.
One can rewrite this model in terms of two symplectic Majorana
spinors in order to exhibit the $R$ symmetry group U(2).} 

The
(3+1)-dimensional Lagrangian then reads
\begin{eqnarray}
  \mathcal L&=&-
  \{\frac{1}{4} F_{\mu\nu}^2+\frac{1}{2}(D_\mu S)^2+\frac{1}{2}(D_\mu P)^2+\frac{1}{2} 
  g^2(S\times P)^2\}\nonumber\\
  \label{eq:L4d} 
  &&-\{
  \bar\psi\gamma^\mu D_\mu\psi +ig\bar\psi (S\times \psi)
+g\bar\psi\gamma_5(P\times\psi)\}.
\end{eqnarray}

We choose
the symmetry-breaking Higgs field as $S^a\equiv A_6^a=v \delta^a_3$
in the trivial sector. The BPS monopoles are
of the form (for $A_0=0$) \cite{Prasad:1975kr}
\bea
A_i^a&=&\epsilon_{aij} {x^j\0g r^2}(1-K(mr)),\\
S^a&=&\delta^a_i{x^i\0g r^2} H(mr),
\eea
with $H=m r \coth(mr)-1$ and $K=mr/\sinh(mr)$,
where $m=gv$ is the mass of the particles that are charged under
the unbroken U(1).
The BPS equation $F_{ij}^a+\epsilon_{ijk}D_k S^a=0$
can be written as a self-duality equation for $F_{MN}$ with $M,N=1,2,3,6$,
and the classical mass is $M_{\rm cl.}=4\pi m/g^2$.

%Classically, this solution saturates the Bogomolnyi bound
%resulting from the susy algebra

The susy algebra for the charges $Q^\alpha=\int j^{0\alpha}d^3x$ with
$j^A={1\02}\Gamma^B \Gamma^C F_{BC} \Gamma^A \lambda$ reads
\be\label{QQPUV}
\{ Q^{\alpha}, \bar Q_{\beta} \}
=-(\gamma^\mu)^\alpha{}_{\beta} P_\mu
+(\gamma_5)^\alpha{}_\beta\, U+i \delta^\alpha_\beta\, V, \quad
\ee
with $\alpha,\beta=1,\ldots,4$.
In the trivial sector $P_\mu$ acts as $\partial_\mu$, and $U$
multiplies the massive fields by $m$, but in the topological sector
$P_\mu$ are covariant translations, and $U$ and $V$ are surface integrals.
The classical monopole solution %(\ref{monsol}) 
saturates the
BPS bound $M^2 \ge |\langle U \rangle|^2+|\langle V \rangle|^2$
by
$|U_{\rm cl.}|=M_{\rm cl.}$, and $V_{\rm cl.}=0$.
{Quantum corrections may change the values of matrix
elements of the charges in (\ref{QQPUV}), but the algebra (\ref{QQPUV})
remains unmodified and the charges  themselves conserved,
i.e., without anomalies, as we discussed above.}

For obtaining the one-loop quantum corrections, one has to
consider quantum fluctuations about the monopole background.
The bosonic fluctuation equations turn out to be simplest
in the background-covariant
Feynman-$R_\xi$ gauge which is obtained
by dimensional reduction of the ordinary 
background-covariant Feynman gauge-fixing term in (5+1) dimensions
$-{1\02}(D_B[\hat A]\,a^B)^2$, where $a^B$ comprises the bosonic fluctuations
and $\hat A^B$ the background fields.
As has been found in Refs.~\cite{Kaul:1984bp,Imbimbo:1985mt},
in this gauge the eigenvalues of the bosonic fluctuation equations
(taking into account Faddeev-Popov fields) and those of the
fermionic fluctuation equations combine such that
one can make use of an index-theorem by Weinberg \cite{Weinberg:1979ma}
to determine the spectral density. This leads to
the following (unregularized!) formula for the one-loop mass correction
\be
M^{(1)}=
{4\pi m_0\0g_0^2} + {\hbar\02} \sum \left( \omega_B-\omega_F \right)
={4\pi m_0\0g_0^2} + {\hbar\02} 
\int\!{d^3k\0 (2\pi)^{3}}
\sqrt{k^2+m^2}\,\rho_M(k^2),
\ee
with $m_0$ and $g_0$ denoting bare quantities and
\be\label{rhoM}
\rho_M(k^2)={-8\pi m\0 k^2 (k^2+m^2)}.
\ee
%As has been observed in Refs.~\cite{Kaul:1984bp,Imbimbo:1985mt},
This expression is logarithmically divergent and is made finite
by combining it with the one-loop renormalization of $g$,
while $m$ does not need to be renormalized \cite{Kaul:1984bp,Imbimbo:1985mt}.

Defining renormalized quantities in the trivial sector and using
the back\-ground-covariant Feynman-$R_{\xi}$ gauge one finds
that tadpole diagrams cancel among themselves. Because of
background-covariance it suffices to formulate a renormalization
condition for one of the two-point functions, and a particularly
simple and natural choice is to renormalize the two-point functions 
of the massless bosons on-shell.
With such a choice, Refs.~\cite{Kaul:1984bp,Imbimbo:1985mt}
came to the conclusion that the counterterms precisely cancel the contribution
from the zero-point energies\footnote{Ref.~\cite{Imbimbo:1985mt} also
considered other renormalization schemes {and other
gauge choices}, where there are finite
remainders {and left the
question of existence of quantum corrections
to the monopole mass to some extent open}. 
In the present paper we shall restrict ourselves to
discussing the above ``minimal'' scheme.}.

However, while \cite{Kaul:1984bp} did not specify the
regularization method used to obtain this result,
Ref.~\cite{Imbimbo:1985mt} regularized by inserting
slightly different
oscillatory factors in the two-point function and in the
integral over the spectral density $\rho_M$, a procedure that is not obviously
self-consistent.

We shall instead use dimensional regularization in a supersymmetry
preserving manner, namely by embedding the (3+1)-dimensional theory
and the BPS monopole in up to one higher dimension (the $x^5$ direction),
where the BPS monopole can be trivially extended into a string-like object.
For the purpose of dimensional regularization of the (3+1)-dimensional
model this is sufficient; it corresponds to trivial Kaluza-Klein
reduction of $x^6$, {and continuous dimensional reduction
from (4+1) to (3+1) dimensions.}
%but it should be noted that going further
%up in dimensionality would eventually lead to a nontrivial embedding
%as the background $S\equiv A_6$ field would require nontrivial
%$x^6$-dependence.

The so dimensionally regularized mode sum can then be written as
\be\label{dimregmodsum}
{\hbar\02} \sum \left( \omega_B-\omega_F \right)
= {\hbar\02} 
\int {d^3k\,d^\epsilon\ell \0 (2\pi)^{3+\epsilon}}
\sqrt{k^2+\ell^2+m^2}\,\rho_M(k^2)
\ee
with $\rho_M$ still given by (\ref{rhoM}).

{At this point we have a choice how to present our results:
we can either show BPS saturation (and its nontrivial ingredients)
at the unrenormalized (but regularized!) level and thus remain
independent of specific renormalization prescriptions, or we
can first renormalize the theory and give well-defined final
results also. Since the `minimal' renormalization scheme
introduced above is the most widely used one and since therein
previous works have obtained null results, we opt for the
latter and just remark that BPS saturation itself as well
as the anomalous contribution that
we shall derive are both independent of
the details of the renormalization procedure.}

Renormalizing the on-shell photon self-energy in background
covariant $R_{\xi=1}$ gauge one obtains
%$Z_m=1$, $m_0=m$, and on-shell photon self-energy in background
%covariant $R_{\xi=1}$ gauge
\be\label{gren}
{1\0g_0^2}={1\0g^2}+4\hbar \int {d^{4+\epsilon}\0(2\pi)^{4+\epsilon}}
{1\0(k_E^2+m^2)^2}
\ee
where the index $\scriptsize E$ in $k_E$ refers to Euclidean signature.

Now,
carrying out the $\ell$-integration in the sum over zero-point energies
(\ref{dimregmodsum}) gives (setting from now on $\hbar=1$)
\be
{1\02}\sum\left( \omega_B-\omega_F \right)
=-{2m\0\pi} {\Gamma(-\2-{\epsilon\02}) \0 (2\pi^\2)^\epsilon \Gamma(-\2)}
\int_0^\infty dk (k^2+m^2)^{-\2+{\epsilon\02}},
\ee
while $3+\epsilon$ integrations in the counterterm in (\ref{gren}) yield
\be
\delta M\equiv 4\pi m(g_0^{-2}-g^{-2})=
16\pi m {\Gamma(\2-{\epsilon\02}) \0 (2\pi^\2)^{3+\epsilon}}
\int_{-\infty}^\infty {dk_4\02\pi} (k_4^2+m^2)^{-\2+{\epsilon\02}}.
\ee

Combining these two expressions we find that there is
a mismatch proportional to $\epsilon$, but $\epsilon$ multiplies
a logarithmically divergent integral, which in dimensional
regularization involves a pole $\epsilon^{-1}$. We therefore
obtain a finite correction of the form
\bea\label{M1M}
M^{(1)}&=&{4\pi m\0g^2}-\epsilon\times
{2m\0\pi}{\Gamma(-\2-{\epsilon\02}) \0 (2\pi^\2)^\epsilon \Gamma(-\2)}
\int_0^\infty dk (k^2+m^2)^{-\2+{\epsilon\02}}\nonumber\\
&=&{4\pi m\0g^2}-{2m\0\pi}+O(\epsilon)
\eea
which because of the fact that it arises as $0\times\infty$ bears the
%hallmark 
fingerprint of an anomaly.

Indeed, as we shall now show, this result is completely analogous
to the case of the $N=1$ susy kink in (1+1) dimensions, where a
nonvanishing quantum correction to the kink mass 
(in a minimal renormalization scheme) is associated
with {an anomalous contribution to }
the central charge (which is scheme-independent;
in a non-minimal renormalization scheme there are also
non-anomalous quantum corrections to the central charge).

%\section{Anomaly in central charge}

In Ref.~\cite{Imbimbo:1985mt} it has been argued that in the
renormalization scheme defined above, the one-loop contributions
to the central charge precisely cancel the contribution from
the counterterm in the classical expression.
In this parti\-cular calculation it turns out that
the cancelling contributions have identical form so that
the regularization methods of Ref.~\cite{Imbimbo:1985mt} can
be used at least self-consistently, and also 
straightforward dimensional regularization
would imply complete cancellations.
The result (\ref{M1M}) would then appear to violate the
Bogomolnyi bound.

However, this is just the situation encountered in the
(1+1)-dimensional susy kink. As we have shown in Ref.~\cite{Rebhan:2002yw},
dimensional regularization gives a
zero result for the correction to the central charge unless the
latter is augmented by the momentum operator in the
extra dimension used to embed the soliton.
This is necessary for manifest supersymmetry, and, indeed,
the extra momentum operator can acquire a nonvanishing expectation value.
As it turns out, the latter is entirely due to nontrivial
contributions from the fermions $\psi={\psi_+\choose\psi_-}$, 
whose fluctuation equations have the form
\bea
L \psi_+ + i(\6_t+\6_5)\psi_- &=& 0,\\
i(\6_t-\6_5)\psi_+ + L^\dagger \psi_-&=&0.
\eea

The fermionic field operator can be written as
\bea
\psi(x)&=&\int\frac{d^{\epsilon}\ell}{(2\pi)^{\epsilon/2}}
\intsum {d^3k\0(2\pi)^{3/2}}{1\0\sqrt{2\omega}}
\biggl\{ a_{kl} e^{-i(\omega t - \ell x^5)}
{\sqrt{\omega-\ell}\; \chi_+ \choose - \sqrt{\omega+\ell} \;\chi_- }\nonumber\\
&&\qquad\qquad\qquad\qquad+b_{kl}^\dagger e^{i (\omega t - \ell x^5)}
{ \sqrt{\omega-\ell} \;\chi_+ \choose  \sqrt{\omega+\ell} \;\chi_- }
\biggr\}
\eea
where $\chi_-={1\0\omega_k}L\chi_+$ and
$\chi_+={1\0\omega_k}L^\dagger\chi_-$ 
with $\omega_k^2=k^2+m^2$, and the
normalization factors $\sqrt{\omega\pm\ell}$ are such that
$L^\dagger L \chi_+ = \omega^2 \chi_+$ and
$L L^\dagger \chi_- = \omega^2 \chi_-$ with $\omega^2=\omega^2_k+\ell^2$.
Because of these normalization factors, one obtains an expression
for the momentum density $\Theta_{05}$
in the extra dimension which has an even
component under reflection in the extra momentum variable $\ell$
\bea
\langle \Theta_{05} \rangle &=&
\int\frac{d^{\epsilon}\ell}{(2\pi)^{\epsilon}}
\int {d^3k\0(2\pi)^3}{\ell\02\omega}
 \left[(\omega-\ell)|\chi_+|^2+(\omega+\ell)|\chi_-|^2\right]\nonumber\\
&=&\int\frac{d^{\epsilon}\ell}{(2\pi)^{\epsilon}}
\int {d^3k\0(2\pi)^3}\frac{\ell^2}{2\omega}(|\chi_-|^2-|\chi_+|^2)
\eea
(omitting zero-mode contributions which do not
contribute in dimensional regularization \cite{Rebhan:2002yw}).

Integration over $x$ then produces the spectral density (\ref{rhoM}) and 
finally yields
\bea\label{deltaZ}
\Delta U_{\rm an}&=&\int d^3x\, \langle \Theta_{05} \rangle =
\int {d^3k\,d^\epsilon\ell \0 (2\pi)^{3+\epsilon}}
{\ell^2 \0 2 \sqrt{k^2+\ell^2+m^2}}\,\rho_M(k^2) \nonumber\\
&=&-4m
\int_0^\infty {dk\02\pi} 
\int {d^\epsilon\ell \0 (2\pi)^{\epsilon}}
{\ell^2 \0 (k^2+m^2)\sqrt{k^2+\ell^2+m^2}}\nonumber\\
&=&-8 {\Gamma(1-{\epsilon\02})\0(4\pi)^{1+{\epsilon\02}}}
{m^{1+\epsilon}\01+\epsilon}
=-{2m\0\pi}+O(\epsilon),
\eea
which is indeed equal to the nonzero mass correction obtained above.

We thus have verified that the BPS bound remains saturated under
quantum corrections, but the quantum corrections to mass and
central charge both contain an anomalous contribution, analogous
to the anomalous contribution to the
central charge of the 1+1 dimensional minimally
supersymmetric kink.

The nontrivial result (\ref{deltaZ}) is in fact in complete accordance
with the low-energy effective action for $N=2$ super-Yang-Mills theory
as obtained by Seiberg and Witten \cite{Seiberg}.\footnote{We are grateful
to Horatiu Nastase for pointing this out to us.}
According to the latter, the low-energy effective action is fully determined
by a prepotential $\mathcal F(A)$, which to one-loop order is
given by
\be\label{F1loop}
\mathcal F_{\rm 1-loop}(A)={i\02\pi}A^2\ln {A^2\0\Lambda^2},
\ee
where $A$ is a chiral superfield
and $\Lambda$ the scale parameter of the theory generated by
dimensional transmutation. 
The value of its scalar component $a$
corresponds in our notation to $gv=m$. In the absence of a $\theta$
parameter, the one-loop renormalized coupling is given by
\be\label{taua}
{4\pi i\0g^2}=\tau(a)={\partial^2 \mathcal F \0 \partial a^2}
={i\0\pi}\left(\ln{a^2\0\Lambda^2}+3\right).
\ee
This definition agrees with the ``minimal'' renormalization scheme
that we have considered above, because the latter involves only the
zero-momentum limit of the two-point function of the massless fields.
For a single magnetic monopole, the central charge 
%$Z \equiv a n_e + a_D n_m$ 
is given by
\be
|U|=a_D={\partial \mathcal F\0\partial a}=
{i\0\pi}a\left(\ln{a^2\0\Lambda^2}+1\right)
={4\pi a\0g^2}-{2a\0\pi},
\ee
and since $a=m$, this exactly agrees with the result
of our direct calculation in (\ref{deltaZ}).

Now, the low-energy effective action associated with (\ref{F1loop})
has been derived from a consistency requirement with the
anomaly of the U(1)$_R$ symmetry of the microscopic theory.
The anomaly in the
{conformal} central charge, which we have identified as being
responsible for the entire
nonzero correction (\ref{deltaZ}), is evidently consistent with
the former. Just as in the case of the minimally supersymmetric kink
in 1+1 dimensions, it constitutes a new anomaly\footnote{%
The possibility of {anomalous contributions to the central charges
of} $N=2$ super-Yang-Mills theories in 4 dimensions 
has most recently also been
noted in \cite{Shifman:2003uh}, however without a calculation
of the coefficients.} 
that had previously
been {missed} in direct calculations \cite{Kaul:1984bp,Imbimbo:1985mt}
of the quantum corrections to %the central charge of 
the $N=2$ monopole using sums over zero-point energies.

{We intend to discuss the details of the anomalous
conformal central-charge current further in a future publication.
We also plan then to consider dimensional regularization by
dimensional reduction starting from 3+1 dimensions (instead of
from 4+1 dimensions as we have done in this paper), which, as we
have shown in Ref.~\cite{Rebhan:2002yw}, locates the anomaly in
an evanescent counterterm.}

\subsection*{Acknowledgments}
We would like to thank S. Mukhi and H. Nastase
for very useful correspondence,
and M. Shifman for drawing our attention to Ref.~\cite{Shifman:2003uh}.
This work has been supported in part by the Austrian Science Foundation
FWF, project no. P15449.
P.v.N. and R.W. gratefully
acknowledge financial support from the International
Schr\"od\-inger Institute for Mathematical Physics, Vienna, Austria.

\small
%\bibliographystyle{elsart-numwot}
%\bibliography{qft,ar,books}

\end{document}